\documentclass[pra,twocolumn,preprintnumbers,amsmath,amssymb,showpacs,
nofootinbib,floatfix]{revtex4}

\usepackage{graphicx,bm,amsmath}

\makeatletter
\def\graphicscale{\twocolumn@sw{0.33}{0.4}}
\def\graphicthreescale{\twocolumn@sw{0.33}{0.4}}
\makeatother

\begin{document}

\title{
The three-dimensional gauge-glass model
}

\author{Vincenzo Alba}
\affiliation{
Max-Planck-Institut f\"{u}r Physik komplexer Systeme, 
D-01187 Dresden, Germany,\\
Scuola Normale Superiore and INFN, I-56126 Pisa, Italy}
\email{vincenzo@pks.mpg.de}
\author{Ettore Vicari}
\affiliation{Dipartimento di Fisica dell'Universit\`a di Pisa
        and INFN, I-56127 Pisa, Italy}
\email{Ettore.Vicari@df.unipi.it}

\begin{abstract}
  
  We investigate the temperature-disorder ($T$-$\sigma$) phase diagram
  of a three-dimensional gauge glass model, which is a cubic-lattice
  nearest-neighbor XY model with quenched random phase shifts $A_{xy}$
  at the bonds, by numerical Monte Carlo simulations.  We consider the
  uncorrelated phase-shift distribution $P(A_{xy})\sim \exp[({\rm
  cos}A_{xy})/\sigma]$, which has the pure XY model and the uniform
  distribution of random shifts as extreme cases at $\sigma=0$ and
  $\sigma\to\infty$ respectively, and which gives rise to equal
  magnetic and overlap correlation functions when $T=\sigma$.
    
  While the high-temperature phase is always paramagnetic, at low temperatures
  there is a ferromagnetic phase for weak disorder (small $\sigma$) and a
  glassy phase at large disorder (large $\sigma$).  These three phases are
  separated by transition lines with different magnetic and glassy critical
  behaviors.  The disorder induced by the random shifts turns out to be
  irrelevant at the paramagnetic-ferromagnetic transition line, where the
  critical behavior belongs to the 3D XY universality class of pure systems;
  disorder gives only rise to very slowly decaying scaling corrections.  The
  glassy critical behavior along the finite-temperature paramagnetic-glassy
  transition line belongs to the gauge-glass universality class, with a quite
  large exponent $\nu=3.2(4)$.  These transition lines meet at a multicritical
  point M, located at $T_M=\sigma_M=0.7840(2)$.  The low-temperature
  ferromagnetic and glassy phases are separated by a third transition line,
  from $M$ down to the $T=0$ axis, which is slightly reentrant.

\end{abstract}



\maketitle


\section{Introduction and summary}
\label{intro}

Spin glass models are simplified, although still quite complex, models
retaining the main features of physical systems which show glassy
phases.  They may be considered as theoretical laboratories where the
combined effects of quenched disorder and frustration can be studied.
Their phase diagrams and critical behaviors provide examples of
possible scenarios which can be used to interpret the experimental
results of complex materials.  For example, Ising-type spin glasses,
such as the $\pm J$ Ising model~\cite{EA-75}, model disordered
uniaxial magnetic materials characterized by random ferromagnetic and
antiferromagnetic short-ranged interactions.  While many theoretical
and numerical works have been devoted to the study of the phase
diagrams and magnetic and glassy critical behaviors of Ising-like spin
glasses, see, e.g., Refs.~\onlinecite{Young-04,KR-03,HPV-08} and
references therein, much less is known about the thermodynamic
properties of spin glass models with continuous symmetries. In the
case of the Heisenberg spin glass there is some evidence for a
finite-temperature glassy transition, but its nature is still debated,
see, e.g., Refs.~\onlinecite{FMPTY-09,VK-09} and references therein.

Another physically interesting spin glass model is the XY model with
random shifts, also known as the gauge-glass model, which is
characterized by a global U(1) symmetry. It has been proposed as a
simplified model of disordered granular superconductors, to describe
vortex-glass
transitions~\cite{SES-84,GK-86,JL-86,HS-90,FTY-91,BFGLV-94,NS-00}.
The phase diagram and critical behaviors of two-dimensional (2D) gauge
glasses have been much investigated, see, e.g.,
Refs.~\onlinecite{APV-10,Korshunov-06} and references therein.  It is
now well established that at weak disorder there is a low-temperature
quasi-long-range order phase separated by a Kosterlitz-Thouless
transition line from the paramagnetic phase; no glassy phase exists at
finite temperature, but a $T=0$ glassy critical behavior at
sufficiently large disorder.  A discussion of the general
temperature-disorder phase diagram of the three-dimensional (3D) gauge
glass can be found in Refs.~\onlinecite{ON-93,Nishimori-book}.  The
ferromagnetic phase of the pure cubic-lattice XY model is expected to
survive at weak disorder~\cite{LNRS-96}, while it should disappear at
large disorder, where a low-temperature glassy phase may exist.
Several numerical works have addressed the existence of a vortex-glass
phase at finite temperature in the case of a uniform random-shift
distribution~\cite{RTYF-91,Gingras-91,CBK-91,KS-97,WY-97,MG-98,
OY-00,KY-01,AK-02,KY-02,KC-05,KWB-07,RD-08}, providing evidence of a
finite-temperature glassy transition.  Apart from these results for
the extreme disordered case, the phase diagram and the critical
behaviors at the different transition lines have not been numerically
investigated yet.  We also mention that experimental results for
vortex glass phases in superconductors have been reported in
Refs.~\onlinecite{KFGKGF-89,KCMESSJ-98,PPKFC-00,SSFPVL-01,LO-02}, but
the experimental scenario for the behavior at the transition does not
appear settled yet.

In this paper we investigate the temperature-disorder phase diagram of the 3D
gauge glass and its critical behaviors along the transition lines which
separate the different phases.

The 3D gauge-glass model is defined by the partition function
\begin{eqnarray}
&&Z(\{A\}) = \int [D\psi] \exp (-{\cal H}/T),\label{RPXY} \\
&&{\cal H} = -\sum_{\langle xy \rangle } {\rm Re} \,\bar\psi_x U_{xy} \psi_y
= - \sum_{\langle xy \rangle} {\rm cos}(\theta_x - \theta_y-A_{xy}),
\nonumber
\end{eqnarray}
where $T$ is the temperature, $\psi_x\equiv e^{i\theta_x}$, $U_{xy}\equiv e^{i
  A_{xy}}$, and the sum runs over the bonds ${\langle xy \rangle }$ of a cubic
lattice.  The phases $A_{xy}$ are uncorrelated quenched random variables with
zero average.  A Gaussian distribution $P_G(A_{xy}) \propto
\exp(-{A_{xy}^2/(2\sigma)})$ is often considered in the studies of 
gauge glasses.  In this paper we consider a slightly different {\em cosine}
distribution
\begin{equation}
P(A_{xy})\propto \exp\left({{\rm cos} A_{xy}\over \sigma}\right).
\label{CRPXYd}
\end{equation}
Analogously to the Gaussian distribution, we recover the pure
cubic-lattice XY model for $\sigma\to 0$ and uniformly distributed
random phase shifts in the limit $\sigma\rightarrow \infty$.  The
cosine distribution is particularly interesting because it lends
itself to some exact relations along the so-called Nishimori (N) line
$T = \sigma$~\cite{ON-93,Nishimori-book}, such as the equality of the
magnetic and overlap correlation functions, which are useful to
identify the multicritical point where the transition lines meet in
the $T$-$\sigma$ phase diagram.

\begin{figure}[tbp]
\includegraphics*[scale=\graphicscale]{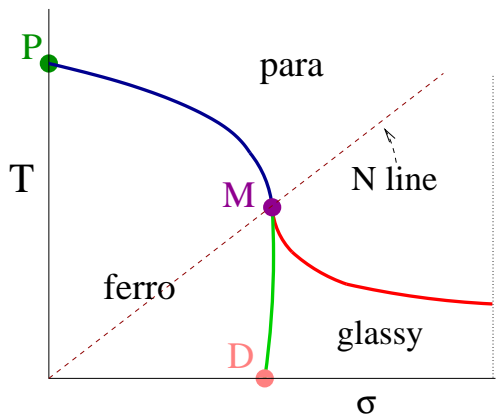}
\caption{Sketch of the temperature-disorder ($T$-$\sigma$)
 phase diagram of the 3D gauge glass.
}
\label{3drpxy}
\end{figure}

\begin{figure}[tbp]
\includegraphics*[scale=\graphicscale]{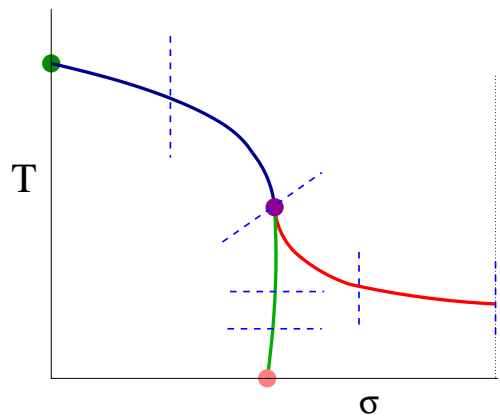}
\caption{The dashed lines sketch the 
$T$ and $\sigma$ values of 
our MC simulations.
}
\label{mcsum}
\end{figure}

Our study of the $T$-$\sigma$ phase diagram, and the magnetic and
glassy critical behaviors along its transition lines, is mostly based
on numerical Monte Carlo (MC) simulations.  Supplementing the
numerical results with renormalization-group (RG) and finite-size
scaling (FSS) analyses, we arrive at the phase diagram sketched in
Fig.~\ref{3drpxy}.  In Fig.~\ref{mcsum} we show where we performed the
MC simulations in the $T$-$\sigma$ plane. Our main results are the
following.

While the high-temperature phase is always paramagnetic, at low
temperatures we have a ferromagnetic phase for weak disorder (small
$\sigma$) and a glassy phase at sufficiently large disorder (large
$\sigma$). These three phases are separated by different transition
lines: a paramagnetic-ferromagnetic (PF) transition line, a
paramagnetic-glassy (PG) transition line and a ferromagnetic-glassy
(FG) transition line, meeting at a multicritical point M located along
the N line.  The phase diagram of 3D gauge glasses presents several
analogies with the temperature-disorder phase diagram of 3D $\pm J$
Ising spin glasses~\cite{KR-03,HPV-08}, where analogous phases 
and transition lines appear.

The 3D gauge glass shows different magnetic and glassy critical behaviors at
the transition lines separating the different phases.  

We argue that the disorder induced by the random shifts is irrelevant at the
PF transition line starting from the pure XY transition point for $\sigma=0$,
at~\cite{CHPV-06} $T_{XY}=2.201842(5)$.  Thus the asymptotic critical behavior
belongs to the 3D XY universality class of pure systems with U(1) symmetry,
characterized by the correlation-length critical exponent~\cite{CHPV-06}
$\nu_{XY}=0.6717(1)$.  Analogously to randomly dilute 3D XY
models~\cite{PV-02}, the disorder induces new scaling corrections which gets
suppressed very slowly, as $O(\xi^{-\omega_{\rm d}})$ where $\xi$ is the
critical length scale and $\omega_{\rm d}=3-2/\nu_{XY}=0.0225(5)$.  A FSS
analysis of MC simulations at $\sigma=0.35$, up to $L=30$, supports this
critical behavior.  The PF transition line ends at the multicritical point M
located at $T_M=\sigma_M\approx 0.784$ along the N line.  For comparison, we
mention that, unlike the gauge-glass model, the quenched disorder is relevant
at the PF transition line of the 3D $\pm J$ Ising models, giving rise to a new
universality class which is the same of the 3D randomly dilute Ising
systems~\cite{HPPV-07-pf,PV-02}.

A low-$T$ glassy phase appears for sufficiently large disorder, i.e.,
$\sigma\gtrsim \sigma_M$ separated by a finite-$T$ PG transition line
from the paramagnetic phase.  A reasonable hypothesis is that the
glassy critical behavior is universal along the PG transition line, up
to $\sigma\to\infty$ (i.e., the model with uniform disorder
distribution), and belongs to the 3D gauge-glass universality
class. This is supported by MC simulations for $\sigma=4/3$ and for
the uniform random-shift distribution, up to lattice size $L=20$,
which provide clear evidences of finite-$T$ transitions in both cases
and the universality of their glassy critical behaviors.  Moreover,
their FSS analyses give $T_c=0.475(10)$ and $T_c=0.46(1)$ respectively
for $\sigma=4/3$ and $\sigma\to\infty$, and the estimates $\nu=3.2(4)$
and $\eta=-0.47(2)$ for the universal exponents describing the
critical overlap correlations.  These critical exponents may be
compared with those of the glassy transition in Ising-like spin
glasses, where~\cite{HPV-08,KKY-06,BCFMPRTTUU-00} $\nu=2.45(15)$ and
$\eta=-0.375(10)$.

The PF and PG transition lines meet at the critical point M along the N line,
see Fig.~\ref{3drpxy}. Actually, M is a multicritical point, characterized by
a magnetic-glassy multicritical behavior with two relevant perturbations (in
the absence of external fields).  The tangents at M 
of the transition lines
are parallel to the $T$ axis.  Indeed, as proved in Ref.~\onlinecite{ON-93},
$\sigma_M$ is an upper bound for the values of $\sigma$ where the
ferromagnetic phase can exist.  A FSS analysis of MC simulations along the N
line, up to $L=20$, locates the point M at $T_{M}=\sigma_{M}=0.7840(2)$, and
provides the estimates $y_1=0.93(3)$ and $y_2=0.56(3)$ for the RG dimensions
of the relevant perturbations, thus a crossover exponent $\phi\equiv
y_1/y_2=1.7(1)$. Moreover, the exponent $\eta$ associated with the spin and
overlap correlation functions is $\eta=-0.121(1)$.  Again, these critical
exponents may be compared with those at the multicritical point of the phase
diagram of the 3D $\pm J$ Ising model where the PF and PG transition lines
meet, which are~\cite{HPPV-07} $y_1=1.02(5)$, $y_2=0.61(2)$, $\phi=1.67(10)$
and $\eta=-0.114(3)$.

A transition line separating the low-$T$ ferromagnetic and glassy phases
starts from the multicritical point M toward the $T=0$ axis, see
Fig.~\ref{3drpxy}.  The order parameter is provided by the magnetic variables
and their correlations, which become effectively paramagnetic in the glassy
phase.  We present FSS analyses of MC simulations at fixed $T<T_M$ up to
$L=12$.  They show that the FG transition line runs almost parallel to the $T$
axis; it is slightly reentrant, consistently with the fact that $\sigma_M$ is
an upper bound for the presence of ferromagnetism.  For example, we find
$\sigma_c=0.777(2)$ at $T=0.376$.  The magnetic critical behavior varying $T$
is compatible with a universal critical behavior along the FG transition line,
with critical exponent $\nu=1.0(1)$.

The main features of the phase diagram and the universality classes of
the magnetic and glassy critical behaviors at the different transition
lines are expected to be largely independent of the detail of the
distribution. For example, they are expected to apply to the case of
random shifts with Gaussian distribution.

The paper is organized as follows.  Sec.~\ref{notations} provides the
definitions of the quantities considered in our work.  In Sec.~\ref{pftl} we
discuss the phase diagram at low disorder and the critical behavior at the PF
transition line.  In Sec.~\ref{pgtl} we focus on the the phase diagram at
large disorder, i.e., large values of $\sigma$, where the low-$T$ phase is
glassy, and study the glassy critical behavior at the PG transition line.  In
Sec.~\ref{mcp} we study the multicritical behavior at the point M of the phase
diagram, where the different transition lines meet.  Finally, in
Sec.~\ref{fgtl} we investigate the FG transition line, which runs from M down
to the $T=0$ axis.

\section{Notations}
\label{notations}

We consider the gauge-glass model (\ref{RPXY}) defined on cubic
lattices of size $L^3$ with periodic boundary conditions.  We define
the magnetic correlation function
\begin{equation}
G(x-y) \equiv [ \langle \bar{\psi}_x \,\psi_y \rangle ]
\label{magcorr}
\end{equation}
and the overlap correlation function
\begin{equation}
G_o(x-y) \equiv [ \langle \bar{q}_x \,q_y \rangle ] \equiv
[\; |\langle \bar{\psi}_x \,\psi_y \rangle|^2 \;], 
\label{overcorr}
\end{equation}
where $q_x$ is the overlap variable defined as 
\begin{equation}
q_x = \bar{\psi}_x^{(1)} \psi_x^{(2)}
\label{qxdef}
\end{equation}
using two copies $\psi^{(1)}_x$ and
$\psi^{(2)}_x$ for the same disorder configuration.  The angular and
square brackets indicate the thermal average and the quenched average
over disorder, respectively.  We define the magnetic and overlap
susceptibilities as
\begin{equation}
\chi\equiv \sum_x G(x), \quad \chi_o\equiv
\sum_x G_o(x), 
\label{chi}
\end{equation}
and the magnetic and overlap second-moment correlation lengths
\begin{eqnarray}
\xi^2 \equiv {\widetilde{G}(0) - 
\widetilde{G}(q_{\rm min}) \over 
          \hat{q}_{\rm min}^2 \widetilde{G}(q_{\rm min}) },
\quad 
\xi_{o}^2 \equiv {\widetilde{G}_o(0) - 
\widetilde{G}_o(q_{\rm min}) \over 
          \hat{q}_{\rm min}^2 \widetilde{G}_o(q_{\rm min}) },
\label{smc}
\end{eqnarray}
where $q_{\rm min} \equiv (2\pi/L,0,0)$, $\hat{q} \equiv 2 \sin q/2$.

We also consider quantities that are invariant under RG
transformations in the critical limit, such as the ratios
\begin{equation}
R_\xi \equiv \xi/L,\quad R^o_\xi\equiv \xi_o/L,
\label{rxi}
\end{equation}
and the cumulants
\begin{eqnarray}
U_{4}  \equiv 
{[\langle |\mu|^4 \rangle ]\over [\langle |\mu|^2 \rangle]^{2}}, 
\quad 
U_{22} \equiv  {[\langle |\mu|^2 \rangle ^2]-
[\langle |\mu|^2 \rangle]^2 \over [\langle |\mu|^2 \rangle ]^2},
\label{cumulants}
\end{eqnarray}
and
\begin{eqnarray}
U_{4}^o  \equiv 
{[\langle |\mu_o|^4 \rangle ]\over [\langle |\mu_o|^2 \rangle]^{2}}, 
\quad 
U_{22}^o \equiv  {[\langle |\mu_o|^2 \rangle ^2]-
[\langle |\mu_o|^2 \rangle]^2 \over [\langle |\mu_o|^2 \rangle ]^2},
\label{cumulantso}
\end{eqnarray}
where $\mu\equiv \sum_x \psi_x$ and $\mu_o\equiv \sum_x q_x$.
Finally, given two replicas of the system with spins $\psi^{(1)}_x$
and $\psi^{(2)}_x$, we consider the quantity~\cite{KY-02,OY-00}
\begin{eqnarray}
I_o= \beta \sqrt{[\langle I^{(1)}\rangle\langle I^{(2)}\rangle]}
\label{irms}
\end{eqnarray}
where $\beta\equiv 1/T$, and
\begin{eqnarray}
I^{(i)}\equiv \frac{1}{L}\sum\limits_{x}{\rm Im}
\,\bar{\psi}_x^{(i)}U_{x\,
  x+\hat{e}_1}\psi_{x+\hat{e}_1}^{(i)} \label{Idef},
\end{eqnarray}
is the derivative of the free energy with respect to a twist
along one direction $\hat{e}_1$ of the lattice.

\section{The paramagnetic-ferromagnetic transition line}
\label{pftl}

The random shifts of the gauge glass model (\ref{RPXY}) vanish when
$\sigma\to 0$, thus recovering the pure cubic-lattice
nearest-neighbour XY model, which undergoes a continuous transition at
$T_{XY}=2.201842(5)$ between the high-$T$ paramagnetic phase and a
low-$T$ ferromagnetic phase with long-range order.  The critical
behavior belongs to the 3D XY universality class~\cite{PV-02},
characterized by the symmetry U(1), which also describes transitions
related to the formation of Bose-Einstein condensates in interacting
quantum particle systems, the superfluid transition in $^4$He,
transitions in easy-plane magnets, etc...  The critical exponents,
which determine the asymptotic behaviors of the critical correlations,
are~\cite{CHPV-06,PV-02} $\nu_{XY}=0.6717(1)$, $\eta_{XY}=0.0381(1)$,
$\alpha_{XY}=2-3\nu_{XY}=-0.0151(3)$, etc...

The low-$T$ ferromagnetic phase is expected to be stable with respect
to the presence of a weak disorder respecting the global U(1)
symmetry, such as that arising in the gauge glass model for nonzero
values of $\sigma$, see also \cite{LNRS-96}, analogously to the 2D
gauge glasses~\cite{APV-09} (where actually we have a low-$T$
quasi-long-range order phase), and the 3D $\pm J$ Ising
model~\cite{HPPV-07-pf} where the disorder does not break the global
Z$_2$ symmetry.  Therefore we expect that a paramagnetic-ferromagnetic
(PF) transition line starts from the pure XY critical point $P\equiv
(\sigma=0,T=T_{XY})$, where the relevant symmetry is still U(1).

\subsection{Irrelevance of the disorder at the PF transition}
\label{irrdis}

The critical behavior along the PF line can be inferred by studying the
relevance of the RG perturbation induced by the random shifts at the pure 3D
XY fixed point.  We argue that, since the global U(1) symmetry is maintained
in the presence of random shifts, the leading RG perturbation induced by the
random shifts gets effectively coupled to the energy density at the PF
transition line, as in the case of randomly-dilute spin
models~\cite{Aharony-76,PV-02}.  This implies that the relevance of the
disorder is related to the sign of the specific-heat exponent of the pure
system~\cite{Harris-74,Aharony-76}.  If it is positive, like the case of 3D
Ising-like models, the disorder provides a relevant perturbation, which
changes the asymptotic critical behavior, giving rise to a new universality
class~\cite{HPPV-07-pf}.  On the other hand, if the specific-heat exponent is
negative, like the case of the 3D XY models where $\alpha_{XY}=-0.0151(3)$
(but also for any 3D model with symmetry O($N$) and $N\ge 2$~\cite{PV-02}),
the disorder is irrelevant in the RG sense, which implies that the critical
behavior belongs to the same universality class of the pure system, and the
asymptotic power-law behaviors remain unchanged.

However, disorder gives rise to scaling corrections which gets suppressed very
slowly and are absent in the pure system.  Analogously to randomly dilute 3D
XY models~\cite{PV-02}, they are $O(\xi^{-\omega_{\rm d}})$ where $\xi$ is the
critical length scale and
\begin{equation}
\omega_{\rm d}=-{\alpha_{XY}\over \nu_{XY}}= 3 - {2\over \nu_{XY}}
=0.0225(5), 
\label{omegad}
\end{equation}
to be compared with the leading scaling correction of pure XY systems
which are $O(\xi^{-\omega})$ with~\cite{CHPV-06,PV-02}
$\omega=0.785(20)$.  

\begin{figure}[tbp]
\includegraphics*[scale=\graphicscale]{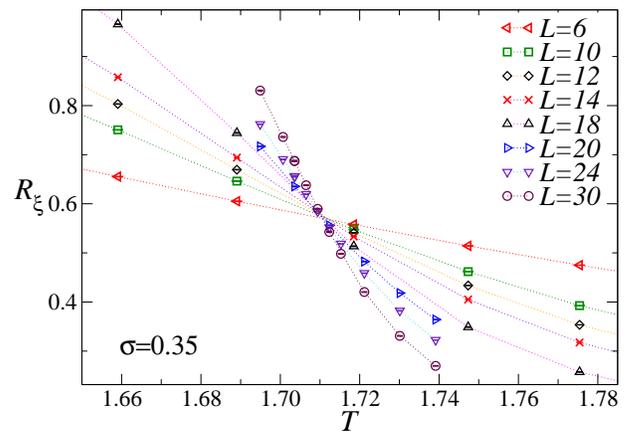}
\caption{
MC data of $R_\xi$ at $\sigma=0.35$.
}
\label{pf_rxi}
\end{figure}

\subsection{Monte Carlo simulations at $\sigma=0.35$}
\label{mcpf}

We support the above scenario by a FSS analysis of MC simulations at
$\sigma=0.35$, up to lattice size $L=30$. In the simulations we use
both Metropolis and microcanonical local updatings.  One single step
of the Monte Carlo update was composed by one Metropolis sweep
followed by 5 microcanonical sweeps.  See Ref.~\onlinecite{APV-10}
for more details. We average over a large number of samples, from
$N_s=16000$ for $L=6$ decreasing to $N_s=3000$ for $L=30$.  The
equilibration is carefully checked by monitoring the MC time
evolution of the observables which we consider.

\begin{figure}[tbp]
\includegraphics*[scale=\graphicscale]{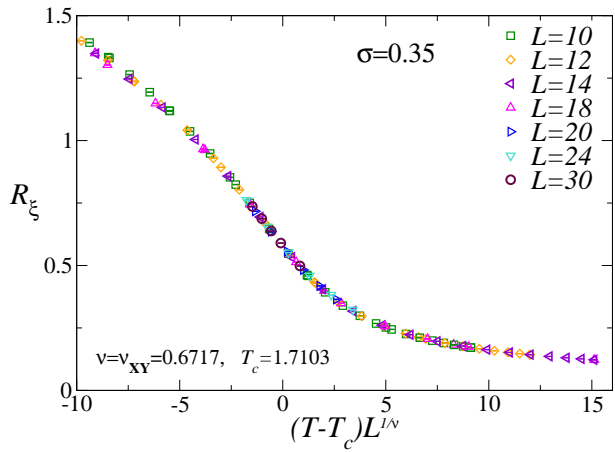}
\caption{
  $R_\xi$ vs $(T-T_c)L^{1/\nu_{XY}}$ with $\nu_{XY}=0.6717$, for
  $\sigma=0.35$.  }
\label{pf_rxi_r}
\end{figure}

The FSS of the RG invariant quantities, such as $R_\xi$, $U_4$, $R^o_{\xi}$,
and $U^o_4$, show a clear evidence of a continuous transition, see, e.g.,
Fig.~\ref{pf_rxi}. A standard FSS analysis gives $T_c=1.7103(3)$ from the
crossing point of their data for different values of $L$, and $\nu=0.68(1)$
from their slope at $T_c$ (using data for $L\ge L_{\rm min}\gtrsim 12$). This
estimate of $\nu$ is in good agreement with the exponent $\nu_{XY}=0.6717(1)$
of the 3D XY universality class. Indeed, Fig.~\ref{pf_rxi_r} shows that a good
collapse of the MC data of $R_\xi$ is achieved by plotting them versus
$(T-T_c)L^{1/\nu_{XY}}$ with $\nu_{XY}=0.6717$, thus supporting the
universality with the 3D XY universality class. Analogous results are obtained
for the other RG invariant quantities. 

\begin{figure}[tbp]
\includegraphics*[scale=\graphicscale]{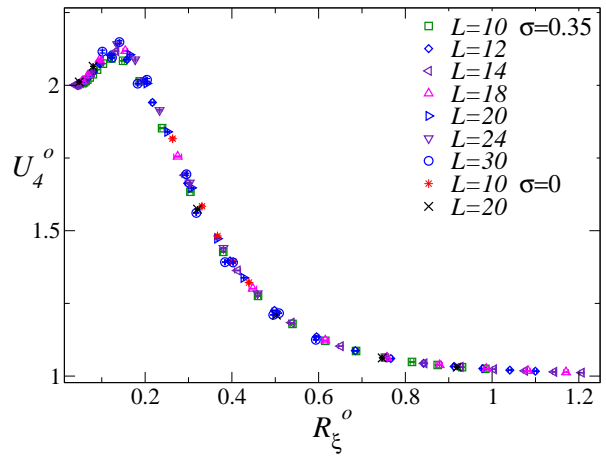}
\caption{
$U_4^o$ vs $R_{\xi}^o$ 
at $\sigma=0.35$ and for the pure XY model ($\sigma=0$).
}
\label{pf_u4o_rxio}
\end{figure}

Given two generic RG invariant quantities $R_1$ and $R_2$, the relation
\begin{equation}
R_1 = f_{R_1}(R_2)
\label{r1fr2}
\end{equation}
is asymptotically universal, i.e., independent of the model within the given
universality class, apart from scaling corrections. This fact provides further
stringent checks of universality.  As an example, Fig.~\ref{pf_u4o_rxio} shows
a plot of $U_4^o$ versus $R_{\xi}^o$ from MC simulations of the gauge glass
with $\sigma=0.35$ and of the pure XY model, i.e., $\sigma=0$. The data appear
to approach a unique universal curve in the large-$L$ limit, providing a
further strong evidence of universality.  Analogous results are obtained for
other combinations of $R_\xi$, $R_\xi^o$, $U_4$ and $U_4^o$.

\begin{figure}[tbp]
\includegraphics*[scale=\graphicscale]{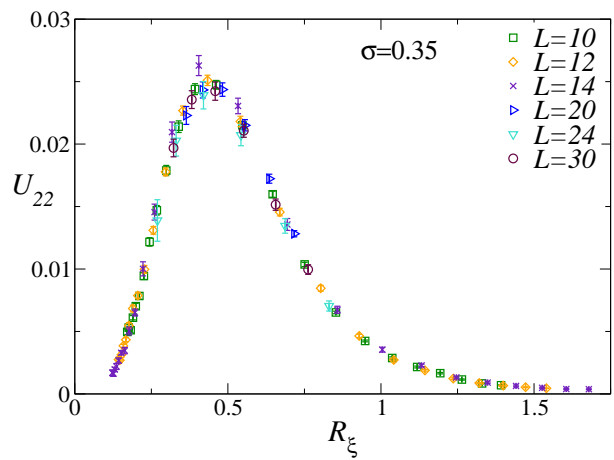}
\caption{
$U_{22}$  vs $R_{\xi}$ at $\sigma=0.35$. 
}
\label{pf_u22_rxi}
\end{figure}

These numerical results provide a strong evidence that the critical behavior
along the PF transition line belongs to the 3D XY universality class of pure
systems.  The above-mentioned slowly decaying scaling corrections are best
observed in the quantity $U_{22}$, cf. Eq.~(\ref{cumulants}),
which is trivially zero in pure systems.
In Fig.~\ref{pf_u22_rxi} we show $U_{22}$ versus $R_{\xi}$ for several values
of $L$.  The data of $U_{22}$ at $\sigma=0.35$ are very small, $U_{22} <
0.03$, but nonzero.  $U_{22}$ is expected to vanish for $L\to\infty$, but very
slowly, as
\begin{equation}
U_{22} \sim L^{-\omega_d}\bar{f}_{U_{22}}(R_\xi), \qquad \omega_d=0.0225(5),
\label{u22eq}
\end{equation}
where $\bar{f}_{22}(R_\xi)$ is a universal function apart from a
trivial overall normalization. Without taking into account
Eq.~(\ref{u22eq}), the data of $U_{22}$ in
Fig.~\ref{pf_rxi_r} might be considered
as an evidence of an unexpected scaling behavior.  However, one can
easily check that they are still compatible with a very slow
suppression in the large-$L$ limit, such as Eq.~(\ref{u22eq}). Indeed,
with increasing the size from $L=10$ to $L=30$, the expected variation
should be just a few per cent, which is within the typical statistical
error of the data in Fig.~(\ref{pf_u22_rxi}).  A check of the
vanishing large-$L$ limit of $U_{22}$ would require much larger
lattices and higher statistics.  A similar situation has been met at
the PF transition line of the 2D $\pm J$ Ising model, where the
$U_{22}$ vanishes logarithmically, as shown in
Ref.~\onlinecite{HPPV-08}.

\section{The paramagnetic-glassy transition line}
\label{pgtl}

In the case of a uniform distribution of random shifts, which is
formally represented by the limit $\sigma\to\infty$, there is already
a good numerical evidence, see, e.g., Refs.~\onlinecite{KY-02,OY-00},
for a finite-temperature PG transition between the high-$T$
paramagnetic phase and a low-$T$ glassy phase.  The glassy phase is
expected to persist when we consider finite large values of $\sigma$;
thus, a PG transition line is expected for sufficiently large values
of $\sigma$. More precisely, it is expected to run from the large
disorder limit down to the multicritical point M, see
Fig.~\ref{3drpxy}, at $T_M=\sigma_M\approx 0.784$, see the next
section.  Our working hypothesis is that the critical behavior along
this PG line is universal, belonging to the 3D gauge-glass
universality class.

\subsection{Monte Carlo simulations}
\label{mcsim}

In order to investigate the glassy critical behavior at large disorder, we
present FSS analyses of MC simulations for $\sigma=4/3$ and for uniformly
distributed phase shifts, formally corresponding to $\sigma\to\infty$, up to
lattice sizes $L=16$ and $L=20$ respectively.  We perform averages
over $10^4$ disorder configurations for all the lattice sizes and temperatures
considered.

In these MC simulations we supplement the local MC updating method,
used at the PF transition line, with the random-exchange or parallel
tempering method, see, e.g, Ref.~\onlinecite{par-temp}, which allows
us to reliably simulate small values of $T$, in particular below the N
line $T = \sigma$.  In the parallel-tempering simulations we consider
$N_T$ systems at the same value of $\sigma$ and at $N_T$ different
temperatures in the range $T_{\rm max} \ge T_i \ge T_{\rm min}$, with
$T_{\rm max}\approx 0.87$ and $T_{\rm min}\approx 0.37$.  The value
$T_{\rm max}$ is chosen so that the thermalization at $T_{\rm max}$ is
sufficiently fast, while the intermediate values $T_i$ are chosen so
that the acceptance probability of the temperature exchange is at
least $5\%$.  Moreover, we require that one of the $T_i$ is along the
N line, i.e., $T_i= \sigma$, where the known exact results allow us to
check the MC code and the thermalization. The thermalization is
further checked by verifying that the averages of the observables
remain stable for all $T_i$ after a sufficiently large number of MC
steps for each disorder realization.  The overlap correlations and the
corresponding $\chi_o$ and $\xi_o$ are measured by performing two
independent runs for each disorder sample.  In the case of observables
requiring the computation of the disorder average of products of
thermal expectations, such as the case of $U_{22}^o$, we use unbiased
estimators as explained in Refs.~\onlinecite{HPPV-07-bias,HPV-08} (a
naive application of the disorder average would introduce a bias, thus
a systematic error).

\subsection{Finite-size scaling analysis}
\label{fss}

\begin{figure}[tbp]
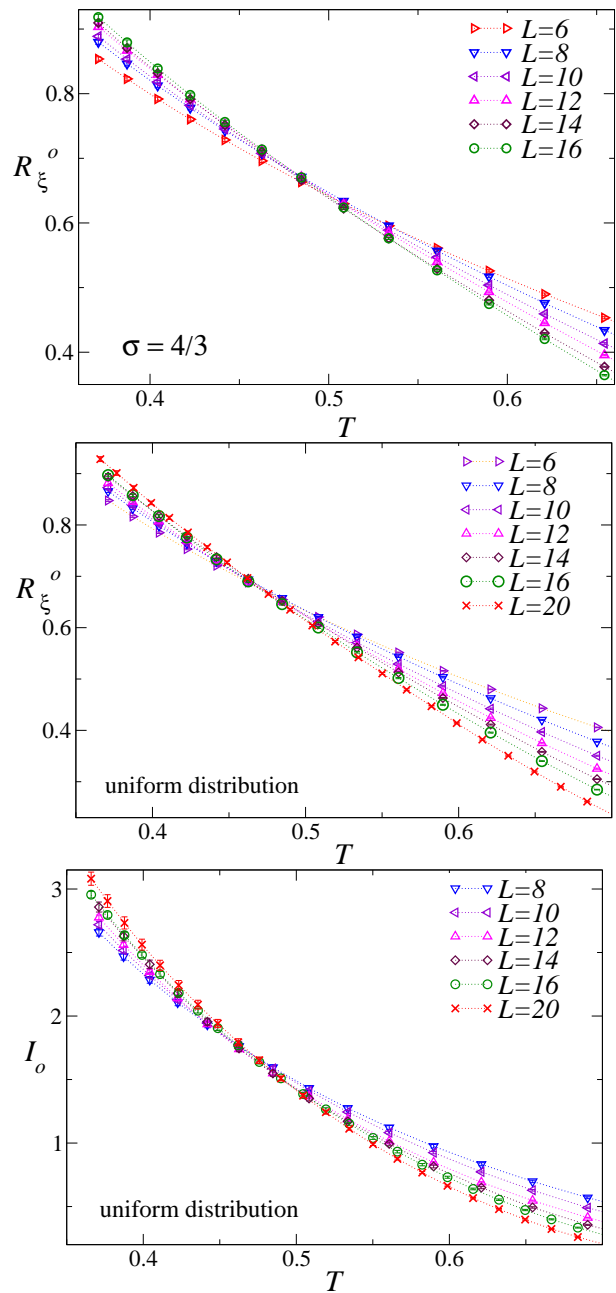

\includegraphics*[scale=\graphicscale]{fig7a.eps}
\includegraphics*[scale=\graphicscale]{fig7b.eps}
\includegraphics*[scale=\graphicscale]{fig7c.eps}
\caption{
  $R_\xi^o$ for $\sigma=4/3$ (above), and $R_\xi^o$ (middle) and $I_o$
  (bottom) for the uniform random-shift distribution.  }
\label{pg_data}
\end{figure}

In Fig.~\ref{pg_data} we show the MC data of $R_\xi^o$ and $I_o$, cf.
Eq.~(\ref{irms}). Their crossing points provide a strong evidence of a
finite-$T$ transition for both $\sigma=4/3$ and $\sigma=\infty$.  A large
exponent $\nu$ is already suggested by the slow rising of the slopes at the
crossing point with increasing $L$.

\begin{figure}[tbp]
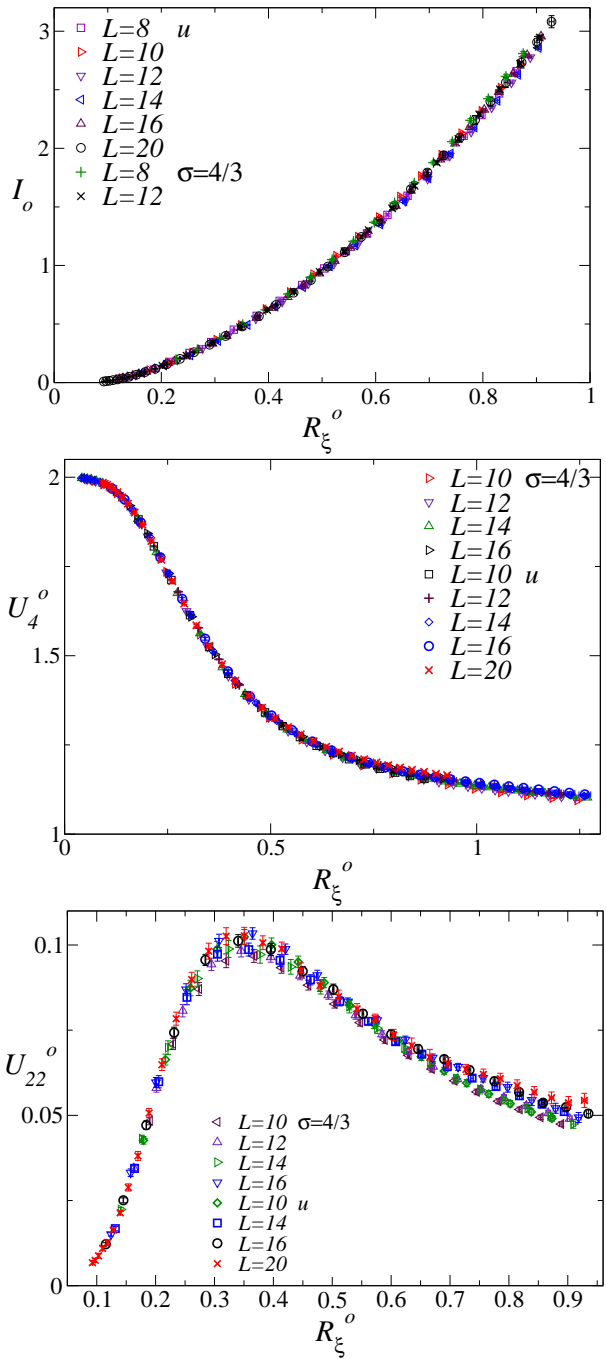

\includegraphics*[scale=\graphicscale]{fig8a.eps}
\includegraphics*[scale=\graphicscale]{fig8b.eps}
\includegraphics*[scale=\graphicscale]{fig8c.eps}
\caption{
  $I_o$ (above), $U_4^o$ (middle) and $U_{22}^o$ (bottom) versus
  $R_\xi^o$ for $\sigma=4/3$ and the uniform distribution
(denoted by ``u'')}
\label{pg_univ}
\end{figure}

To beghin with, we address the universality issue, i.e., whether the
transitions at $\sigma=4/3$ and $\sigma=\infty$ have the same
universal critical behavior.  For this purpose, in Fig.~\ref{pg_univ}
we plot data of $I_o$, $U_4^o$ and $U_{22}^o$ versus $R_\xi^o$ at
$\sigma=4/3$ and $\sigma=\infty$. They appear to converge toward the
same universal large-$L$ limit, providing a strong evidence of
universality. These results support our working hypothesis that the
glassy critical behavior is univeral alsong the PG transition line
from large disorder to the multicritical point M.

The FSS behavior of the RG invariant quantities allows us to estimate
$T_c$ and $\nu$:
\begin{eqnarray}
&&R(T,L) = f(u_tL^{1/\nu})= \label{rfit} \\
&&=R^* + c_1 (T-T_c) L^{1/\nu} + c_2 (T-T_c)^2 L^{2/\nu} + ...,
\nonumber
\end{eqnarray}
where $u_t$ is the temperature scaling field, $u_t\sim T-T_c$, $R$
indicates the generic RG invariant quantity, such as $R_\xi^o$ and
$I_o$, and we neglect scaling-correction terms. We check the stability
of the fits by increasing the minimum size $L_{\rm min}$ of the data
used in the fit, varying the range of values of $T$ around $T_c$ [we
use self-consistent windows around $T_c$ limiting the value of
$(T-T_c)L^{1/\nu}$, which corresponds to limiting the range of values
of any $R$ around $R^*$], and the number of terms in Eq.~(\ref{rfit}).
We use the comparison of the results using different quantities as a
check of the relevance of the neglected scaling corrections.  
Some results of fits of the data of $R_\xi$ and $I_o$, using the
Ansatz (\ref{rfit}), are reported in Table~\ref{tabfits}.

\begin{table}
\caption{ Results of fits to $R^* + \sum_{i=1}^n c_i (T-T_c)^i
L^{i/\nu}$ of the data of $R_\xi^o$ and $I_o$, 
with respect to variations of the range of $T$ (taking data for
$|R_\xi^o-R_\xi^{o*}|\le \Delta R_\xi^o$ to have a self-consistent 
scaling cut, with $R_\xi^{o*}\approx 0.7$), the number $n$
of terms in the fit Ansatz, and $L_{\rm min}$ which is the minimum
size allowed by the data.  }
\label{tabfits}
\begin{ruledtabular}
\begin{tabular}{cclcclllc}
\multicolumn{1}{c}{$\sigma$}&
\multicolumn{1}{c}{$R$}&
\multicolumn{1}{c}{$\Delta R_\xi$}&
\multicolumn{1}{c}{$n$}&
\multicolumn{1}{c}{$L_{\rm min}$}&
\multicolumn{1}{c}{$T_c$}&
\multicolumn{1}{c}{$1/\nu$}&
\multicolumn{1}{c}{$R^*$}&
\multicolumn{1}{c}{$\chi^2/{\rm dof}$}\\
\colrule
$\infty$ & $R_\xi^o$ & 0.07 & 1 &  8 &  0.467(2) & 0.27(3) & 0.684(5) & 3.1 \\
        &&      &   & 10 &  0.477(3) & 0.29(4) & 0.665(7) & 2.9 \\
        &&      &   & 12 &  0.458(4) & 0.31(7) & 0.703(9) & 1.8 \\
        && 0.1  & 1 &  8 &  0.465(2) & 0.27(2) & 0.689(4) & 3.6 \\
        &&      &   & 10 &  0.475(2) & 0.29(2) & 0.669(5) & 3.5 \\
        &&      &   & 12 &  0.455(4) & 0.29(4) & 0.712(9) & 2.1 \\
        &&      &   & 14 &  0.461(5) & 0.33(6) & 0.698(11)& 1.8 \\
        && 0.1  & 2 &  8 &  0.466(2) & 0.29(8) & 0.687(3) & 3.2 \\
        && 0.2  & 2 &  6 &  0.463(1) & 0.31(2) & 0.690(2) & 3.3 \\
        &&      &   &  8 &  0.460(1) & 0.32(2) & 0.698(2) & 3.0 \\
        &&      &   &  10&  0.469(2) & 0.32(2) & 0.680(3) & 2.8 \\
        &&      &   & 12 &  0.450(3) & 0.29(3) & 0.721(6) & 1.6 \\ 
        &&      &   & 14 &  0.456(4) & 0.31(4) & 0.708(9) & 1.3 \\ 
&$I_o$   & 0.1  & 1 &  8 &  0.467(5) & 0.26(4) & 1.74(5) & 1.4 \\
        &&      &   & 10 &  0.472(6) & 0.27(5) & 1.69(5) & 1.5 \\
        &&      &   & 12 &  0.488(7) & 0.33(7) & 1.53(7) & 0.8 \\
        && 0.3  & 2 & 8  &  0.463(2) & 0.38(12) & 1.77(2) & 1.6 \\\hline
$4/3$ & $R_\xi^o$ & 0.1 & 1 &  6 &  0.486(2) & 0.31(2) & 0.669(4) & 2.7 \\
    & &     &   &  8 &  0.477(2) & 0.31(2) & 0.685(5) & 1.5 \\
        &&      &   & 10 &  0.478(5) & 0.28(5) & 0.683(10) & 1.8 \\
        && 0.1  & 2 &  8 &  0.476(3) & 0.31(11) & 0.686(4) & 1.0 \\
        && 0.2  & 2 &  6 &  0.479(1) & 0.33(2) & 0.682(2) & 3.1 \\
        &&      &   &  8 &  0.471(2) & 0.33(2) & 0.695(3) & 1.4 \\
        &&      &   & 10 &  0.471(3) & 0.31(3) & 0.695(6) & 1.6 \\ 
\end{tabular}
\end{ruledtabular}
\end{table}

In the case of the uniform distribution we have data up to $L=20$.
The fits of the data of $R_\xi^o$ are reasonably stable. We note some
oscillations in the estimates of $T_c$, which may reflect the
difficulty to estimate $T_c$ when the critical exponent $\nu$ is
large.  From these results one may get the estimates $T_c=0.46(1)$,
$1/\nu=0.31(4)$ and $R_\xi^{o*} = 0.70(2)$, where errors take also
into account the stability with respect to changes of $L_{\rm min}$,
from $L_{\rm min}=6$ to $L_{\rm min}=12$, and the range of $T$ around
the crossing point. The results using $I_o$ are definitely consistent,
although they appear less precise (also because they correspond to
less statistics, we started later collecting data for $I_o$).  They
suggest the estimates $T_c=0.47(2)$, $1/\nu=0.30(5)$ and $I_o^{*} =
1.7(1)$.  The fits of the data for the cumulants $U_4^o$ and
$U_{22}^o$ do not provide sufficiently stable results, but they scale
consistently as shown by Fig.~\ref{pg_univ}.

In the case of the distribution with $\sigma=4/3$, whose data are up
to $L=16$, the results appear more stable.  We obtain $T_c=0.475(10)$,
$1/\nu=0.32(4)$ and $R_\xi^{o*} = 0.69(1)$ from the fits of the data
of $R_\xi^o$.  For $\sigma=4/3$ we have only few data for $I_o$, see
Fig.~\ref{pg_univ}, which do not allow us to obtain an independent
estimate of $\nu$.

\begin{figure}[tbp]
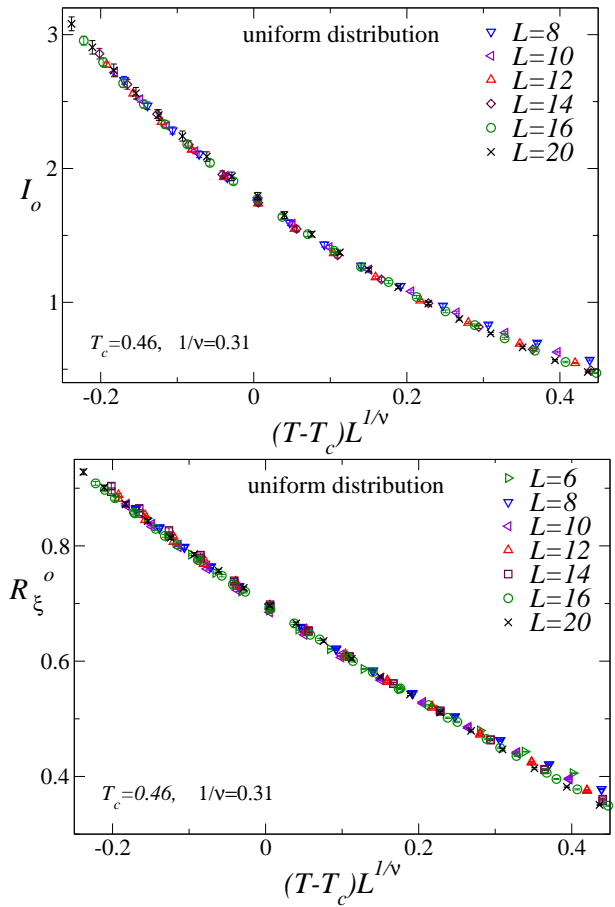

\includegraphics*[scale=\graphicscale]{fig9a.eps}
\includegraphics*[scale=\graphicscale]{fig9b.eps}
\caption{
  $I_o$ (above) and $R_\xi^o$ (bottom) versus $(T-T_c)L^{1/\nu}$ for
  the uniform distribution, with $T_c=0.46$ and $1/\nu=0.31$.  }
\label{pg_chenu}
\end{figure}

The above results for $\nu$ are consistent, thus, in agreement with
universality. We consider
\begin{equation}
1/\nu=0.31(4), \quad \nu = 3.2(4),
\label{nuggest}
\end{equation}
as our final estimate.  Fig.~\ref{pg_chenu} shows the collapse of the
data of $R_\xi^o$ and $I_o$ for the gauge glass with uniform
distribution, when they are plotted versus $(T-T_c)L^{1/\nu}$, with
$T_c=0.46$ and $1/\nu=0.31$.  Note that $\nu$ is quite large, but
still comparable with the value $\nu=2.45(15)$ of the glassy
transition in 3D Ising-like spin
glasses~\cite{HPV-08,KKY-06,BCFMPRTTUU-00}.  We should also say that
the accuracy and the relatively small lattice size of the available
data do not allow us a robust control of the neglected
scaling-corrections, as it was achieved for the Ising-like spin glass
models~\cite{HPV-08-l}.  Therefore, further numerical work is required
to substantiate the above results.

We compute the exponent $\eta$ associated with the overlap correlation
(\ref{overcorr}), by analyzing the data of the overlap susceptibility
$\chi_o$, which is expected to behave as~\cite{HPV-08}
\begin{equation}
\chi_o = \overline{u}_h^2 L^{2-\eta_o} f(u_t L^{1/\nu})~.
\label{chiobeh}
\end{equation}
Here $\overline{u}_h$ is related to the external overlap scaling field $u_h$
associated with the overlap variables by $u_h=h\overline{u}_h(T)+ O(h^2)$.
Thus, neglecting nonanalytic scaling corrections, we fit the data of $\chi_o$
to
\begin{equation}
\ln \chi_o = a \ln L + b_0 + b_1 (T-T_c) + ...
+ c_1 (T-T_c)L^{1/\nu} + ...
\label{acfiteta}
\end{equation}
fixing the values of $T_c$ and $\nu$ as obtained above. We obtain
\begin{equation}
\eta=-0.47(2)
\label{etag}
\end{equation}
for the uniform distribution and $\eta=-0.46(2)$ for the distribution
with $\sigma=4/3$, where the error takes also into account the
uncertainty on $T_c$ and $\nu$.

Finally, we compare our results with earlier numerical works for the
gauge-glass model with uniformly distributed random shifts. Earlier estimates
of $T_c$, i.e., $T_c =0.47(3)$ from Ref.~\onlinecite{OY-00} and $T_c =0.48(2)$
from Ref.~\onlinecite{KWB-07}, are consistent with ours.  Our result
$\nu=3.2(4)$ is significantly larger than the estimates obtained by earlier
FSS analyses, $\nu=1.3(4)$ from Ref.~\onlinecite{RTYF-91}, $\nu=1.39(20)$ from
Ref.~\onlinecite{OY-00}, and $\nu=1.62(20)$ Ref.~\onlinecite{KWB-07}.  These
smaller values of $\nu$ appear excluded by our FSS analyses.  This discrepancy
may be explained by the small lattice sizes, up to $L=12$, of their data.  We
also mention that our estimate is quite larger than the experimental estimate
at the vortex-glass transition in the (K,Ba)BiO$_3$ cubic supercondutor
reported in Ref.~\onlinecite{KCMESSJ-98}, i.e., $\nu=1.0(2)$, which has been
often compared with the glassy transition of the gauge glass with uniform
distribution.  Finally, our estimate (\ref{etag}) of $\eta$ agrees with the
result $\eta=-0.47(7)$ reported in Ref.~\onlinecite{OY-00}.

\section{The multicritical point along the N line}
\label{mcp}

In the case of the cosine distribution (\ref{CRPXYd}), the N line
plays an important role in the phase diagram, because it marks the
crossover between the magnetic-dominated region and the
disorder-dominated one.  We conjecture that the PF and PG transition
lines meet at a multicritical point M, which coincides with the
critical point along the N line, analogously to the phase diagram of
the 3D $\pm J$ Ising model~\cite{HPV-08,HPPV-07,Nishimori-book}.  This
is also suggested by the fact that the location of the critical point
along the N line provides a bound on the disorder parameter $\sigma$
where a ferromagnetic phase can exist~\cite{ON-93}.  We also mention
that the critical point along the N line shows a multicritical
behavior also in the 2D $\pm J$ Ising model and 2D gauge glass, see,
e.g., Refs.~\onlinecite{HPPV-08-mc,PHP-06,APV-10}, even though these models
do not have a low-$T$ glassy phase, thus a finite-$T$ PG transition
line.  We shall support the above scenario by a FSS of MC simulations
of the 3D gauge glass along the N line.

\subsection{The Nishimori line}
\label{nishline}

The cosine distribution (\ref{CRPXYd}) lends itself to some exact calculations
along the N line~\cite{ON-93,Nishimori-book}. For example, the energy density
is~\cite{ON-93} $E = -3{I_1(1/T)/I_0(1/T)}$ when $T=\sigma$, where $I_i$ are
modified Bessel functions.  Moreover, the spin and overlap correlation
functions are equal:
\begin{eqnarray}
[\langle \bar{\psi}_x \psi_y \rangle ] = 
[|\langle \bar{\psi}_x \psi_y \rangle|^2 ].
\label{cnl}
\end{eqnarray}
Along the N line we also have $R_\xi=R^o_\xi$ and $U_4=U^o_4$. 

As proved in Ref.~\cite{ON-93}, the critical value $\sigma_M$ of
$\sigma$ along the N line is an upper bound for the values of $\sigma$
where the ferromagnetic long-range order can exist.  Therefore, at the
critical point M$\;\equiv (\sigma_M,T_M)$ the tangent to the
transition line limiting the ferromagnetic phase must be parallel to
the $T$ axis; moreover, the critical value $\sigma_D$ where the
ferromagnetism disappears at $T=0$ must satisfy $\sigma_D\le\sigma_M$.

The N line $T=\sigma$ is also characterized by an extension of the replica
symmetry. The quenched average over disorder, which implies the average of the
free energy, can be formally reconstructed by introducing $n$ replicas of the
system, taking the limit $n\to 0$ after the disorder average.  We
write the disorder average over the partition function of $n$ replicas as
\begin{eqnarray}
&&[Z^{(n)}(A)] = 
\int [DA] \exp[(1/\sigma)\sum_{\langle xy \rangle}  
{\rm cos} A_{xy}]
\times \label{zn}\\
&& \int \prod_n [D\psi^{(n)}]
\exp[(1/T)\sum_{\langle xy \rangle} 
 {\rm cos}(\theta_x^{(n)}-\theta_y^{(n)}- A_{xy})]
\nonumber
\end{eqnarray}
Let us perform the gauge transformation
\begin{equation}
\psi_x\to e^{i\varphi_x} \psi_x,
\quad 
U_{xy} \to e^{-i\varphi_x} U_{xy} e^{i\varphi_y}.
\label{gtra}
\end{equation}
We obtain
\begin{eqnarray}
&&[Z^{(n)}(A)] = 
\int [DA] \exp[(1/\sigma)\sum_{\langle xy \rangle}  
{\rm cos}(\varphi_x-\varphi_y- A_{xy})]
\times \nonumber \\
&& \int \prod_n [D\psi^{(n)}] 
\exp[(1/T)\sum_{\langle xy \rangle} 
 {\rm cos}(\theta_x^{(n)}-\theta_y^{(n)}- A_{xy})].
\end{eqnarray}
Since a further integration with respect to $\phi_x\equiv
e^{i\varphi_x}$ gives only  rise to a  
trivial constant factor, we have that
the gauge variables $\phi_x$ correspond to another identical replica
when $T=\sigma$, extending the original replica symmetry.

\subsection{Scaling behavior at the multicritical point}
\label{scbeh}

As a working hypothesis, we assume that the critical point along the N
line is a multicritical point, analogously to the case of the 3D $\pm
J$ Ising model~\cite{HPPV-07}. We derive some predictions which are
then verified by a FSS analysis of numerical MC simulations.

In the absence of external fields, the critical behavior at the
multicritical point M is characterized by two relevant RG operators. The
singular part of the free energy averaged over disorder in a volume of
size $L$ can be written as
\begin{equation}
F_{\rm sing}(T,\sigma,L) = L^{-d} f(u_1 L^{y_1}, u_2 L^{y_2}, 
\{u_i L^{y_i}\}),
\label{freeen}
\end{equation} 
with $i\ge 3$,
where $y_1>y_2>0$, $y_i<0$ for $i\ge 3$, $u_i$ are the corresponding scaling
fields, and $u_1 = u_2 = 0$ at M.  In the
infinite-volume limit and neglecting subleading corrections, 
we have
\begin{equation}
F_{\rm sing}(T,\sigma) = |u_2|^{d/y_2} f_\pm (u_1 |u_2|^{-\phi})
\label{freeen2}
\end{equation} 
around M, with $\phi=y_1/y_2>1$,
where the functions $f_\pm(x)$ apply to the parameter regions in which 
$\pm u_2 > 0$. Close to M, all transition lines correspond to 
constant values of the product $u_1 |u_2|^{-\phi}$ and thus,
since $\phi > 1$, they are tangent to the line $u_1 = 0$.

The relevant scaling field $u_1$ and $u_2$ can be inferred by using
the following facts: (i) since $\sigma_M$ is an upper bound for the
values of $\sigma$ where the ferromagnetic phase can
exist, the transition line at M must be parallel to the
$T$ axis; (ii) the condition $T=\sigma$ at the N line is RG invariant,
because it is protected by the extension of the replica symmetry, as
shown in Sec.~\ref{nishline}.  We therefore have
\begin{equation}
u_1 = \sigma-\sigma_M + ...,
\label{u1scalfields}
\end{equation}
where the dots indicate nonlinear corrections, which are quadratic in
$\Delta\sigma\equiv \sigma-\sigma_M$ and $\Delta T\equiv T-T_M$, so that the
line $u_1=0$ runs parallel to the $T$ axis at M.  Moreover, we choose
\begin{equation}
u_2=T-\sigma,
\label{scalfields}
\end{equation}  
so that the N line corresponds to $u_2=0$.  

These results give rise to the following predictions for the FSS behavior
around M. Let us consider a RG invariant quantity
$R$, such as $R_\xi$, $U_4$, $U_{22}$, defined in Sec.~\ref{notations}.  In
general, in the FSS limit $R$ obeys the scaling law
\begin{equation}
R = {\cal R}(u_1 L^{y_1}, u_2 L^{y_2}, \{u_i L^{y_i}\}),\quad i\ge 3.
\label{scalR}
\end{equation}
Neglecting the scaling corrections which vanish in the limit $L\to
\infty$, we expect
\begin{equation}
R = R^* + b_{11} u_1 L^{y_1} + b_{21} u_2 L^{y_2} + \ldots.
\label{scalR1}
\end{equation}
Along the N line, the
scaling field $u_2$ vanishes, so that we can write
\begin{equation}
R_N = R^* + b_{11} u_1 L^{y_1} + b_{21} u_1^2 L^{2y_1}+ \ldots,
\label{RGinvsca}
\end{equation}
where the subscript $N$ indicates that $R$ is restricted to the N line.
Differentiating Eq.~(\ref{scalR1}) with respect to $\beta\equiv 1/T$, we
obtain
\begin{equation}
R' = b_{11} u'_1 L^{y_1} + b_{21} u'_2 L^{y_2} + \cdots 
\label{RGinvdsca1}
\end{equation}
If Eq.~(\ref{u1scalfields}) holds, then $u'_1=O(T-T_M)$, so that
the leading behavior along the N line is
\begin{equation}
R'_N = b_{21} u'_2 L^{y_2} + \cdots 
\label{RGinvdsca2}
\end{equation}
The magnetic susceptibility along the N line behaves as
\begin{equation}
\chi_N = e L^{2-\eta}\left( 1 + e_1 u_1 L^{y_1} + \cdots\right).
\label{RGinvchi}
\end{equation}
Note that there is only one $\eta$ exponent which characterizes the critical
behavior of both the magnetic and overlap correlation functions, since they
are equal along the N line.

\subsection{MC results}
\label{mcmcres}

In the following we present a FSS analysis of MC simulations along the
N line.  The MC algorithm was a mixture of standard Metropolis and
microcanonical updates, as in the MC simulations at the PF transition
line reported in Sec.~\ref{pftl}.  In order to locate the
multicritical point we simulated several temperatures ranging from
$T=0.773$ to $T=0.797$, for lattice sizes $6\le L \le 20$.  The number
of disorder configurations ranged from $4\cdot 10^5$ to $2\cdot 10^7$
for the largest lattices. This large statistics was necessary to
achieve a convincing evidence of the multicritical nature of M.

\begin{figure}[tbp]
\includegraphics*[scale=\graphicscale]{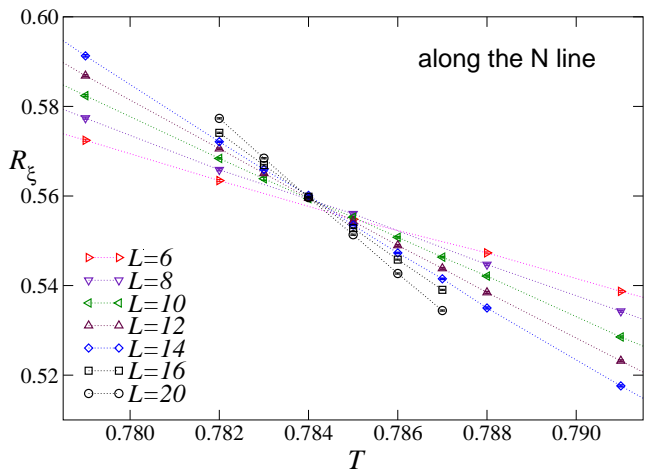}
\caption{
$R_\xi$ along the N line $T=\sigma$.}
\label{mc_rxi}
\end{figure}

The MC data of $R_\xi$ are shown in Fig.~\ref{mc_rxi}.  There is clearly a
crossing point at $T \approx 0.784$.  Analogous results are obtained from
$U_4$ and $U_{22}$.  In order to estimate $T_M$ and $y_1$, we fit the RG
invariant $R$ to
\begin{equation}
R = R^* + a_1 (T - T_M) L^{y_1} + a_2 (T - T_M)^2 L^{2y_1} + ....
\label{fitfo}
\end{equation}
Note that this functional form relies on the property that $u_2 = 0$
along the N line.  Otherwise, an additional term of the form $(T -
T_M) L^{y_2}$ should be added. We also neglect scaling corrections
which are $O(L^{y_3})$ with $y_3 < 0$.  
We check their relevance by comparing 
results from the analyses of different quantities.

Our best estimates are
\begin{eqnarray}
&&T_M=\sigma_M = 0.7840(2), \label{estTM} \\
&&y_1 = 0.93(3), \label{esty1}
\end{eqnarray}
and also $R_\xi^*=0.5594(4)$, $U_4^*=1.226(1)$ and
$U_{22}^*=0.128(4)$.  The errors take into account the stability of
the results with respect to the minimum size $L_{\rm min}$ allowed in
the fits (typically from $L_{\rm min}=6$ to $L_{\rm min}=12$), the
range of values of $T$ around $T_c$ [we use again self-consistent
windows around $T_c$ limiting the values of $(T-T_M)L^{y_1}$], the
number of terms in Eq.~(\ref{fitfo}).  Scaling corrections are
apparently small. We find some evidence of scaling corrections only in
the analysis of $U_{22}$, but they appear to decay quite fast,
suggesting a relatively large scaling correction exponent, i.e.,
$y_3\approx -2$.

The derivative $R'$ with respect to $\beta$
can be estimated by computing appropriate connected
correlations of $R$ and the Hamiltonian in the MC simulations along
the N line.  According to the multicritical scenario outlined above, $R'$
is expected to behave as $L^{y_2}$ at $T_M$ with $y_2<y_1$.  In order
to determine $y_2$, we fit the data to 
\begin{equation}
\ln R' = a + y_2 \ln L + b (T - T_M) L^{y_1},
\end{equation}
keeping $T_M=0.7840$ and $y_1 = 0.93$ fixed.
We obtain
\begin{equation}
y_2 = 0.56(3)
\label{y2est}
\end{equation}
from $R_\xi'$ (the error includes the uncertainty on $T_M$ and $y_1$), while
the data of $U_4'$ are not sufficiently precise to provide a stable result.
The above estimate of $y_2$ nicely support the multicritical scenario, since
it shows that $y_2<y_1$.  Therefore the crossover exponent, cf.
Eq.~(\ref{freeen2}), is 
\begin{equation}
\phi = {y_1\over y_2} = 1.7(1).
\end{equation}

We determine the exponent $\eta$ from the FSS of the ratio $Z \equiv
\chi/\xi^2\sim L^{-\eta}$ at the critical point. We fit its MC data along the
N line to
\begin{equation}
\ln Z = a - \eta \ln L + b (T - T_M) L^{y_1} + c(T-T_M)
\end{equation}
where the last term takes into account possible analytic terms coming from the
scaling fields~\cite{HPV-08}, analogously to Eq.~(\ref{chiobeh}).  We obtain
\begin{equation}
\eta = -0.121(1)
\label{etaest}
\end{equation}

It is worth noting that the above estimates of the multicritical
exponents are quite close to those found for the 3D $\pm J$ Ising
model, at the multicritical point along its N line,
where~\cite{HPPV-07} $y_1=1.02(5)$, $y_2=0.61(2)$, $\phi=1.67(10)$ and
$\eta=-0.114(3)$.

\section{The ferromagnetic-glassy transition line}
\label{fgtl}

The FG transition line runs from the multicritical point down to
$T=0$, with $\sigma_c\le \sigma_M$, since $\sigma_M$ provides a bound
for the region where the magnetic order can exist.  The order parameter
along this transition is provided by the magnetic variables and their
correlations.

Ref.~\onlinecite{ON-93} argues that this transition line runs
parallel to the $T$ axis, thus $\sigma_c=\sigma_M$ for $T<T_M$.  An
analogous prediction for the 2D $\pm J$ Ising model turns out to fail,
although it provides a good approximation, because the low-$T$
transition line where ferromagnetism disappears turns out to be almost
parallel and only slightly reentrant, see, e.g.,
Refs.~\onlinecite{PPV-09,PHP-06} and references therein.

\begin{figure}[tbp]
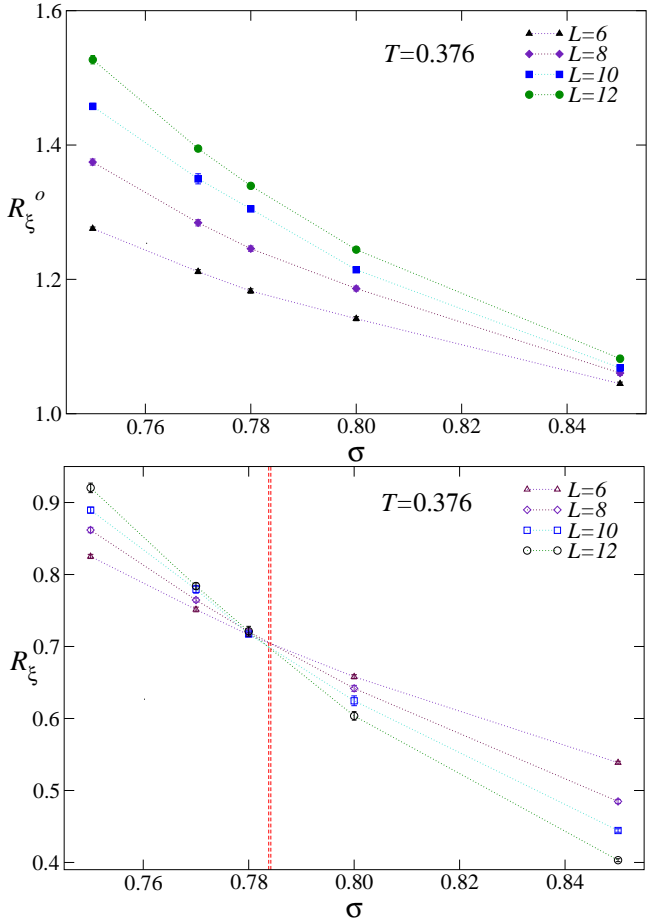

\includegraphics*[scale=\graphicscale]{fig11a.eps}
\includegraphics*[scale=\graphicscale]{fig11b.eps}
\caption{
  $R_\xi$ (below) and $R_\xi^o$ (above) at $T=0.376$ across the FG transition
  line.  The vertical dashed lines show the estimate of $T_M$, i.e.,
  $T_M=0.7480(2)$, with its uncertainty. }
\label{fg_rxi}
\end{figure}

We investigate this issue by numerical MC simulations up to lattice
sizes $L=12$, using the same MC method employed at the glassy
transitions.  Using random-exchange MC simulations, we collect data
down to $T=0.376$ for several values of $\sigma$ in the range $0.75\le
\sigma \le 0.85$.  Our results are obtained by averaging over a large
number of of disorder configurations: $N_s=16000,\,8000,\,6000,\,4000$
respectively for $L=6,\,8,\,10,\,12$.  Fig.~\ref{fg_rxi} shows the
data of $R_\xi$ and $R_\xi^o$ at $T=0.376$.  The set of data of
$R_\xi$ for different lattice sizes show a crossing point, confirming
the existence of FG transition.  On the contrary, the data of
$R_\xi^o$ do not show crossings, which may reflect the fact that 
such transition separates two {\em ordered} phases with respect to
the overlap variables.

Note that the crossing points of the $R_\xi$ data appear to cluster at a value
of $\sigma$ which is slightly smaller than $\sigma_M=0.7840(2)$.  Indeed,
fitting them to
\begin{equation}
R_\xi = R_\xi^* + a_1 (\sigma - \sigma_c) L^{1/\nu} + 
a_2 (\sigma - \sigma_c)^2 L^{2/\nu} + ...,
\label{fitfgrxi}
\end{equation}
we obtain
\begin{equation}
\sigma_c = 0.777(2), \quad \nu=1.0(1),
\label{fgexp}
\end{equation}
and $R_\xi^*=0.73(1)$. These estimates should be taken with some caution, in
particular that of $\nu$, due to the small size of the available lattices,
which does not allow us to perform stringent checks of stability.  The
analysis of the data of $U_4$ gives consistent results, but less precise.  We
also mention that an analogous FSS analysis of the data at a larger
temperature $T=0.437$, but still smaller than $T_M$, gives
$\sigma_c=0.782(2)$, $\nu=1.1(1)$ and $R_\xi^*=0.71(1)$, which support the
universality of the magnetic critical behavior along the FG transition line.

According to the above results, the critical values of the disorder parameter
at $T<T_M$ are very close but smaller than $\sigma_M=0.7840(2)$, indicating a
slight reentrant transition line.

\acknowledgements The MC simulations were performed at the INFN Pisa
GRID DATA center, using also the cluster CSN4.  In total, they took
approximately 100 years of CPU time on a single core of a recent
standard commercial processor.

\end{document}